\documentclass[10pt,journal,compsoc]{IEEEtran}
\ifCLASSOPTIONcompsoc
  \usepackage[nocompress]{cite}
\else
  \usepackage{cite}
\fi
\ifCLASSINFOpdf
\else
\fi
\usepackage{algorithmic}
\usepackage{balance}
\usepackage{graphicx}
\usepackage{listings}
\usepackage[justification=centering]{caption}
\usepackage{booktabs} 
\usepackage{adjustbox}
\usepackage{booktabs} 
\usepackage{amsmath}
\usepackage{array, booktabs, makecell}
\usepackage{flexisym}
\usepackage{comment}
\usepackage[hyphens]{url}
\usepackage{framed}
\usepackage{tablefootnote}
\usepackage{textcomp}
\PassOptionsToPackage{dvipsnames}{xcolor}
\usepackage{xcolor}
\definecolor{dkgreen}{rgb}{0,0.6,0}
\definecolor{gray}{rgb}{0.5,0.5,0.5}
\definecolor{mauve}{rgb}{0.58,0,0.82}
\usepackage{fancybox}
\newcommand{\N}[1]{{\color[rgb]{0.0, 0.0, 0.75}#1}} 
\newcommand{\hypobox}[1]{\begin{center}%
    \noindent\thicklines\setlength{\fboxsep}{8pt}%
    \cornersize{0.2}\Ovalbox{\begin{minipage}{3.1in}%
        \textit{#1}\end{minipage}} \end{center}}

\makeatletter
\begin{document}
%
\title{A Methodology for Analyzing Uptake of Software Technologies Among Developers}

\author{Yuxing~Ma,~\IEEEmembership{StudentMember,~IEEE,}
        Audris~Mockus,~\IEEEmembership{Fellow,~IEEE,}
        Beth Milhollin,
        Russel~Zaretzki, Randy~Bradley and Bogdan~Bichescu
\IEEEcompsocitemizethanks{\IEEEcompsocthanksitem Y. Ma is with the Department
of Electrical Engineering and Computer Science, 
University of Tennessee, Knoxville.\protect\\
E-mail: yma28@vols.utk.edu
\IEEEcompsocthanksitem A. Mockus is with the Department
of Electrical Engineering and Computer Science, 
University of Tennessee, Knoxville.\protect\\
E-mail: see http://mockus.org
\IEEEcompsocthanksitem B. Milhollin, R. Zaretzki, R. Bradley and B. Bichescu are with the Department
of Business Analytics and Statistics, University of Tennessee.}
}
\IEEEtitleabstractindextext{%
\begin{abstract}
Motivation: The question of what combination of attributes drives the adoption of a particular software technology is critical to developers. It determines both those technologies that receive wide support from the community and those which may be abandoned, thus rendering developers' investments worthless. 
Aim and Context: We model software technology adoption by developers and provide insights on specific technology attributes that are associated with better visibility among alternative technologies. Thus, our findings 
have practical value for developers seeking to increase the adoption rate 
of their products. 
Approach: We leverage social contagion theory and statistical modeling to identify, define, and test empirically measures that are likely to affect software adoption. More specifically, we leverage a large collection of open source version control repositories (containing over 4 billion unique versions) to construct a software dependency chain for a specific set of R language source-code files. We formulate logistic regression models, where developers'  software library choices are modeled, to investigate the combination of technological 
attributes that drive adoption among competing data frame (a core concept for a data science languages) implementations in the R language: {\tt tidy} and {\tt data.table}. 
To describe each technology, we quantify key project attributes that might affect adoption (e.g., response times 
to raised issues, overall deployments, number of open defects, knowledge base) and 
also characteristics of developers making the selection (performance needs, scale, and their social network). 
Results: We find that a quick response to raised issues, a larger number of overall deployments, and a larger number of high-quality StackExchange questions are associated with higher adoption.  Decision makers tend to adopt the technology that is closer to them in the technical dependency network and in author collaborations networks while meeting their performance needs.
Future work: We hope that our methodology encompassing social contagion that captures both rational and irrational preferences and the elucidation of key measures from large collections of version control data provides a general path toward increasing visibility, driving better informed decisions, and producing more sustainable and widely adopted software.
\end{abstract}

}

\maketitle

\IEEEdisplaynontitleabstractindextext

%
\IEEEpeerreviewmaketitle

\IEEEraisesectionheading{\section{Introduction}\label{sec:introduction}}
\IEEEPARstart{O}{pen} source has revolutionized software development by creating and
enabling both a culture and practice of reuse, where developers can
leverage a massive number of software languages, frameworks,
libraries, and tools (we refer to these as software technologies) to
implement their ideas. Open source allows
developers, by building on the existing work of others, to focus on their own
innovation~\cite{von2001innovation,VONKROGH20031217,doi:10.1093/oxrep/17.2.248,west2006patterns}, potentially reducing lead times and effort. This approach, however,
is not absent of risks.  For example, if a particular technology
chosen by a developer is later supplanted by another, incompatible
technology, the support for the supplanted technology is likely to
diminish. Reductions in support for the supplanted technology result
in increased effort on the part of the developer to either provide
fixes upstream or to create workarounds in their
software. Furthermore, the value of the developer's creation to new
downstream projects may diminish in favor of the now more
popular alternative package.  As a consequence, both the
importance of a developer's product and their reputation may
suffer.  To remediate these two risks, developers must understand
how attributes of their software products may be perceived among
potential and actual downstream adopters (consumers of the technology), 
especially in relation to alternative, competing technologies these
adopters may have.    
It is natural, therefore, to adopt the position that open source
software development should be investigated from a supply chain
perspective, which also pertains to distributed decision and supply
networks among different stakeholders. We refer to the collection of
developers and groups (software projects) producing updates (patches and new versions) of the source code as a Software Supply Chain (SSC)~\cite{holdsworth1995software,farbey1999exploiting}. The upstream and
downstream links from project to project are represented by the
source code dependencies, sharing of the source code, and by the
contributions via patches, issues, and exchange of
information. While the product adoption in supply chains has been
well studied
~\cite{huang2002product,kalish1985new,russell2004people,christopher2004mitigating}, little is known or understood about how developers choose what
components to use in their own software projects. 

As a complex dynamical system, every player in the open source
ecosystem may have their specific set of preferences or biases, which can affect the ultimate outcome of wide (or narrow)
adoption and/or entire abandonment of formerly popular technologies. 
These decisions
are not only based on technical merit but the availability and accessibility of relevant information along with the tastes of consumers(adopters). 
Furthermore, these SSC networks may severely limit developer choices
at the particular point in time when they need to make decisions on
which components or technologies to use based on what components they 
are aware of and how much time or inclination they have to investigate
the relative merits of the possible choices. This suggest the potentially 
strong influence 
of default choice well documented in behavioural economics. 
Hence, in contrast to common conventions,  we should not simply
model the preferences of individual developers but must also take
into account the complexity of the supply networks and their specific position within them. 

We want to address this major gap in knowledge empirically by using
a very large data source comprising version control data of millions
of software projects.  Our methodology involves using this data to
construct software supply chain networks, identifying software
technology choices,  
theorizing about factors that characterize the developer  and the
technologies they chose, and finally fitting and interpreting the
models for specific technology choices and, thus,
characterizing the implicit primary factors (social, behavioural, and rational) they may use to make their decision.

Despite the practical and theoretical importance of the question
how developers make technology choices, the extant literature 
does not offer theoretical guidance on this subject.
We, therefore, leverage social contagion theory, which has been effective, among other things, 
in clarifying key aspects of organizational adoption of technology~\cite{angst2010social,SAMADI2016263}. Social contagion theory mimics models of the spread of contagious diseases but apply them in the behavioral/social context instead of the physiological one. The first key concept is exposure or how widespread the infectious agent in the population. In our case the agent is a specific package and the population is the entire collection of FLOSS repositories. The exposure is critical in epidemiology because without the exposure a disease can not spread. This brings us to 
\newline
\textbf{RQ1:} Does the exposure to a technology, such as the number of FLOSS repositories in existence, the rate at which new repositories are adopting this technology, or the number of high-quality questions on StackExchange affect the 
decisions of the developers to adopt that technology?

The second key concept is infectiousness: a highly virulent agent is more likely to spread in a population. We deal with packages or technologies (groups of packages), so in our case
we'd like to establish:\newline
\textbf{RQ2:}
Will extremely attractive package (with few open issues, short response times to issues or pull requests, heavy activity and many authors), be more likely to be adopted? 

The final concept is proximity: some infectious agents may not survive the travel through air or physical barriers, thus halting their spread. In our case, the distance from a developer to a package is physical, but it may be represented the technological constraints 
(lack of compatibility with other packages the developer already uses), need for certain performance characteristics, or a social distance to collaborators who are working with other developers already exposed to the package or a related technology. Hence:\newline
\textbf{RQ3:}
Will proximity of a developer or a project to a package increase the rate of adoption? 
More specifically, \textbf{RQ3a:} will the proximity of a developer to a related technology used by a developer increase the chances of adoption;  \textbf{RQ3b:} will the proximity of a developer to collaborators who already use a package or a related technology increase the chances of adoption?; \textbf{RQ3c:} will the performance requirements of the project a developer is working on increase the chances of adoption of a package that has the desired performance attibute?. 


To answer RQs, we need to collect data on the actual choices made by
developers and operationalize key theory-based measures, we need an
to reconstruct the states of all public software projects that may choose the
technology under study. For example, for a project that chooses Package A in January 2014, we need to establish how many other projects have used before that date (exposure), what average response time to issues the project had at that time (infectiousness), and what actions the developer making the choice to add the dependence had prior to that point in time, including her social network, technology network, etc. 

To exemplify the proposed methodology, we investigate rapidly
growing data-science software ecosystem centered around the {\tt R}
language. One of the key
technology choices in this area are the data structures used to
store data (in the data-science sense). R has two major competing
technologies implemented in packages {\tt data.table} and {\tt
  tidy}. (a more detailed introduction of these two packages is given in
Sec.~\ref{Case study})

Our research provides several theoretical and practical innovations.
From the theoretical standpoint, the novelty of our contribution first lies in introducing social 
contagion theory that provides first-principles based methods to
construct hypotheses and to determine measures that should affect
technology adoption. The second novelty is the context in which we investigate 
technology choices, i.e., a complete SSC~\cite{chhajed2005software,ellison2010supply}, not restricted to a set of projects or
ecosystems. Third, we use regression models to understand how macro trends
at the scale of the entire SSC emerge from actual decisions the
individual developers make to select a specific software technology.
More specifically, as a result of contextualizing social contagion theory through 
SSCs, our approach provides novel
measures, such as proximity in a dependency network and authorship
network, questions and answers with high quality in Q\&A,
performance needs, and total deployments, that strongly affect the
spread of technology and that were not used in prior work on library migration.

From the practical standpoint, our contribution 
consists of proposing a method to explain and
predict the spread of technologies, to suggest which
technologies are more likely to spread in the future, and suggest
steps that developers could take to make the technologies they
produce more popular.  Developers can, therefore, reduce risks by
choosing technology that is likely to be widely adopted.  The
supporters of open source software could use such information to focus on and properly 
allocate limited resources on projects
that either need help or are likely to become a popular
infrastructure.  In essence, our approach unveils previously unknown 
critical aspects of  technology spread and, through that, makes developers, organizations, 
and communities more effective. 

In Sec.~\ref{s:theory} we introduce the diffusion of innovation,
social contagion, and the application of choice models. In
Sec.~\ref{s:ssc}, we describe the dataset and how we operationalize
software supply chain. Choice model
theory and our candidate technology are introduced in Sec.~\ref{ss:choice} and Sec.~\ref{s:sctc} respectively. In Sec.~\ref{Case study},
operationalization of attributes of choice model is illustrated. 
Sec.~\ref{s:results} describes and interprets the result of applying
the choice model. Related work is discussed in
Sec.~\ref{s:related} and major limitations are
considered in Sec.~\ref{s:limitations}. We summarize our
conclusions and contribution in Sec.~\ref{s:conclusions}. 
\section{Conceptual background}\label{s:theory}
We draw on methodologies from a diverse set disciplines.  The phenomena we are investigating is often called adoption~\cite{bass1969new} or diffusion of innovation~\cite{rogers1995innovation}. 
Both theoretical approaches model how products or ideas become
popular or get abandoned.  We would like to fit such models and, in
order to do so, 
find relevant set of predictors that have theoretical
justification. Fichman~\cite{fichman2004going} considered how
internal factors  
such as resources and organization predict innovations in commercial
enterprises, and DiMaggio~\cite{10.2307/2095101}
included the factor of  
environment as well. The adopters of the technology may influence
non-adopters over time. Angst \textit{et
  al.}~\cite{angst2010social}, use the concept of  
social contagion~\cite{Burt:social}, which consists of observation, information
transmission, and learning to study spread of electronic health
records. 
These concepts are familiar to any open source developer. More
specifically, in addition to purely social contagion, we also have  
technical dependencies that act as strong constraints on developer actions. The signaling theory applied for social coding 
platforms~\cite{dabbish2012social,tsay2014influence} provides some
specific guidance as to what may motivate developers to chose one
project over another. Many of the actions developers take on GitHub
are focused on building or maintaining their reputation, hence they
pay a particular attention to measures such as activity, numbers of
participants, or ``stars"\footnote{placing a star on a GitHub
  repository allows a developer to keep track of projects they find
  interesting and to discover similar projects in their news
  feed.}. 

The basic premise of social contagion theory is that developers may observe the actions and decisions of others, communicate them, and learn to emulate them over time. This premise implies that 
groups and individuals  who are in social and spatial proximity to prior adopters are more susceptible to the influence of prior adopters of technology. This susceptibility (synonymous with potency or infectiousness of influence) is likely to result in an increased likelihood to adopt the same technology~\cite{angst2010social}.  Notice, that the susceptible to influence 
of prior adopters represents non-rational behaviour. Rational behaviour would require developer to choose the best technology irrespective of social influences. It may also represent cognitive bias of the default choice. The developer may not know about the alternatives if their social or technical networks do not 
present them with an encounter with alternatives. This would represent the irrational bias toward default choice. 
These precursors of spread, if measured and calibrated with the actual level of technology spread, would provide the relative importance of each factor in driving the adoption and provide the understanding to help developers choose technologies wisely and provide hints on how to make their own technology more widely adopted. Fortunately, the mathematical adoption models have been developed and refined over time. A variation of multinomial regression models also called choice models\cite{mcfadden1973conditional} can be used
to describe the behavior of a decision maker given a set of alternatives. 
Choice models have been used successfully in
the fields of marketing \cite{choicemodelKa,HAUSMAN19951,Talluri:2004:RMU:989113.989136,Tim2004} and economics
\cite{mcfadden2000mixed,berry1994estimating,Kenneth1981} to understand how consumers make
choices. Adapting and applying these regression models to technology adoption,
we focus on a developer, or more precisely, a software project as a
decision maker. The actual decision is operationalized as the first 
among the alternative technologies that a project in a commit modifying one of the files within a repository.
As with the social contagion theory, two types
of predictors can be included: properties of the choice (i.e., the technology) and
properties of a decision maker (i.e., the project or individual
developer).  

Equiped with this theoretical and modeling framework, we set out
to address RQ1 and RQ2 by empirically characterizing the spread of software technology through analysis of a very large collection (VLC) of version control data introduced in~\cite{msr09} and curated by OSCAR project~\footnote{bitbucket.org/swsc/overview}. We refer to it as OSCAR-VLC. According to the curators, OSCAR-VLC approximates the entirety of
public version control and includes major forges such as 
GitHub, BitBucket, GitLab, Bioconductor, SourceForge, now defunct
Googlecode, and many others and currently contained over 46M projects at the time of analysys. 
OSCAR-VLC production involves discovering~\cite{ma2016crowdsourcing} and cloning the projects, extracting Git objects from each repository, and then 
storing these objects in a scalable key-value database.

OSCAR-VLC is used to construct the SSC~\cite{greenfield2003software,levy2003poisoning} by
determining dependencies among software projects and developers,
then by characterizing these projects  according to their technical
characteristics and supply chains. The social contagion and
signaling theories allow us to select meaningful measures for the
decision makers and for their choices. (We present our measures in Sec.\ref{ss:measures}) 
\section{Constructing software supply chains}\label{s:ssc}
\begin{figure}[htbp]
\centerline{\includegraphics[width=8cm,height=6cm,keepaspectratio]{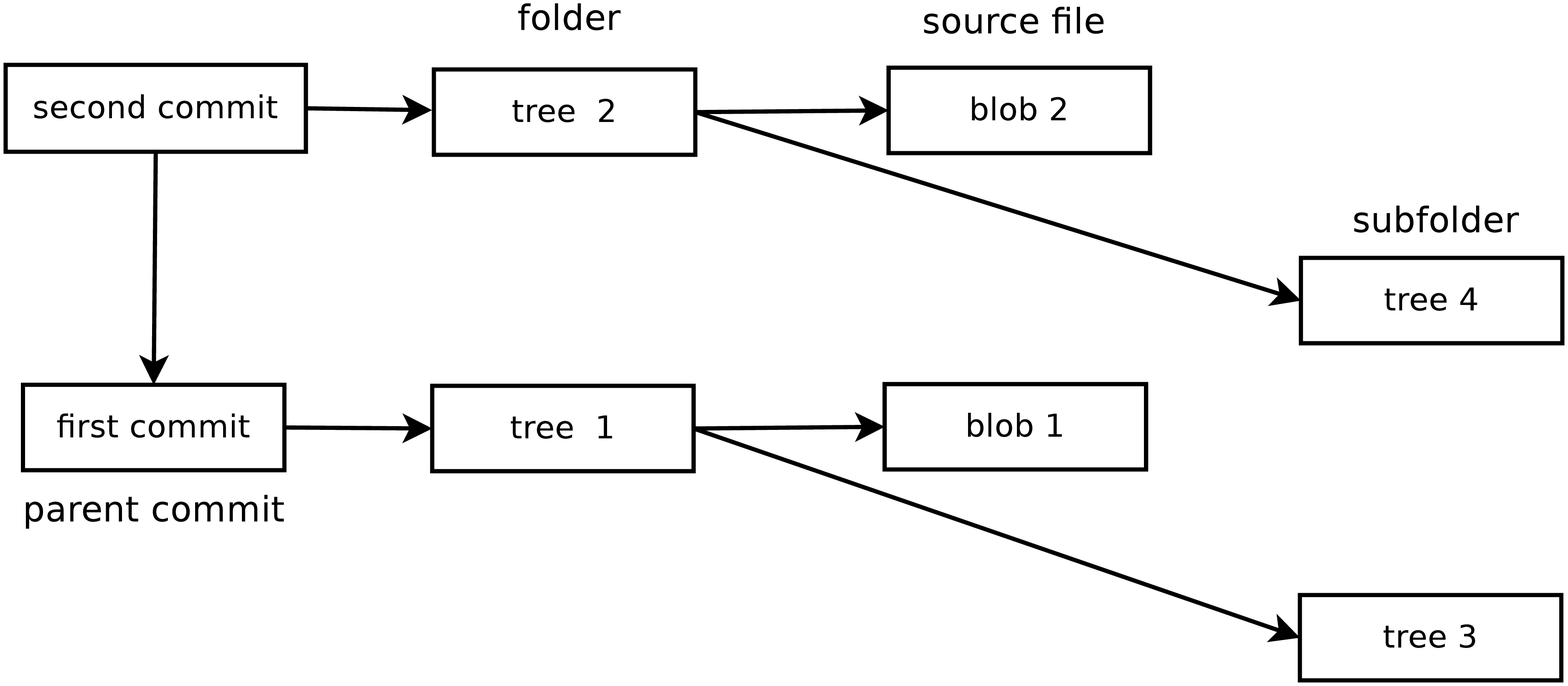}}
\caption{Git objects graph}
\label{fig2}
\end{figure}

Source code changes made in software projects are recorded in a VCS
(version control system) used by the software project. Many of the
projects are using git as their version control system, sometimes
with historic data imported from SVN or other VCS used in the
past. Code changes are typically organized into commits that make
changes to one or more source code files.
Internally, the Git database has three primary types of objects: commits, trees, and blobs~\cite{chacon2014pro}. Each object is represented by its sha1 value that can be used to find its content. The content of a blob object is the content of a  specific version of a file. The content of a tree object is, essentially, a folder in a file system represented by the list of sha1s for the blobs and the trees (subfolders) contained in it. 
A commit contains the sha1 for the corresponding tree, a list of parent commit sha1s, an author string, a committer string, a commit timestamp, and the commit message. 
Fig.~\ref{fig2} illustrates relationships among objects described above.

We utilize all Git objects  (1.1 billion commits and 4 billion of blobs and trees) 
from OSCAR-VLC to construct the relevant supply chain, social, and adoption measures. 
For our analysis we create mappings among these objects and their attributes, e.g., filename to associated blobs.  
\subsection{Software supply chain}
In traditional supply chains~\cite{christopher1992,chopra2007supply,John2011,Hau1997,buurman2002supply}, the networks include material,
financial and information relationships. Similar concepts can be
operationalized in the software domain, with  
developers or projects representing the nodes and information
transfer or static dependencies among projects representing links. 
Based on the characteristics of software domain, especially the open source community, and the ability to measure various attributes  
relevant to technology adoption, we consider two different types of
network relationships: dependency networks, and authorship
networks. 
\subsection{Measuring the dependency network}\label{s:discovery}
While many types of static dependencies exist, here we focus on
explicit specification of the dependency in the source code. For
example, `import' statements in Java or Python, `use' statements in
Perl, `include' statements in C, or, as is the case for our study,
`library' statements for the R language.

We analyze the entire set of 4 billion blobs existing in the database at the time of the analysis using following steps: 
\begin{enumerate}
\item Use file to commit map to obtain a list of commits (and files) for all R language files by looking for the filename extension `.[rR]\$'
\item Use filename to blob map to obtain the content for all versions of the R-language files obtained in Step 1
\item Analyze the resulting set of blobs to find a statement indicating an install or a use of a package:
  \begin{itemize}
  \item \verb/install\.packages\(.*"PACKAGE".*\)/
  \item \verb/library\(.*[\"']*?PACKAGE[\"']*?.*\)/
  \item \verb/require\(.*[\"']*?PACKAGE[\"']*?.*\)/
  \end{itemize}
\item Use blob to commit map to obtain all commits that produced these blobs and then use the commit to determine the date that the blob was created
\item Use commit to project map to gather all projects that installed the relevant set of packages
\end{enumerate}

\begin{figure}[htbp]
\centerline{\includegraphics[width=12cm,height=6cm,keepaspectratio]{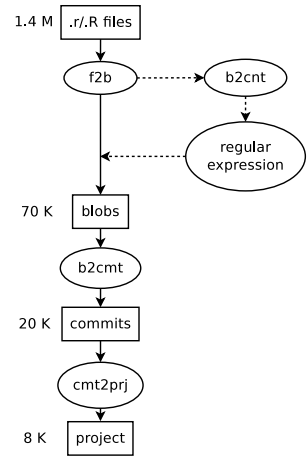}}
\caption{Project discovery}
\label{fig}
\end{figure}

These steps are illustrated in a flowchart in Fig.~\ref{fig}.
In Fig.~\ref{fig}, the rectangular boxes represent inputs and outputs, and ovals represent maps or dictionaries we utilized in this study. f2b stands for filename-To-blob map, b2cnt stands for blob-To-content map, b2cmt stands for blob-To-commit map, and cmt2prj for commit-To-project map. The number on the left side represents the unique number of corresponding objects.

A similar approach can be applied to other languages with suitable modification in the dependency extraction procedures, since different package managers or different languages might require alternative approaches to identify dependencies or the instances of use.

In addition to dependencies, we also need to obtain measures that describe various aspects of social relationships among developers because the theories of adoption, such as social contagion theory we employ, need measures of information flows among individuals as an important factor driving the rate of adoption.
\subsection{Measuring the authorship network}
The authorship network can be viewed as the process of developers working with other developers either by implicitly learning skills from other's 
contribution (source code) or by explicitly communicating through emails or discussion platforms. Here we focus on the former mode of communication 
since the bulk of direct communication may be private. We consider two types of links among developers. A weak link exists between a pair of developers if they commit in at least one project that is common between them and a strong link exists if they change at least one file in common.
\subsection{StackExchange}
StackExchange is a popular question answer website related to programming.
When people search for information there they may notice answers that suggest the use of either {\tt tidy} or {\tt data.table} (discussion about choosing these packages is in Sec.~\ref{s:sctc}) and, consequently, might be inclined to incorporate one of these packages into their own code. 
The latest (2017-12-08) StackExchange data dump including 57GB of posts was imported into MongoDB, out of which 6k questions (excluding answers) were 
found to be related to either {\tt data.table} or {\tt tidy} by searching for these two terms in the title or the content of the post. 
We operationalize two measures: one counts the total number of posts while another measure counts only questions that have a score above 20 to gauge the amount of 
high-quality content that is likely to be referred to from search engines.
\subsection{The Choice Model}\label{ss:choice}
The choice set (set of alternatives) needs to exhibit three
characteristics to be able to fit a discrete choice model. First,
the alternatives need to be \textit{mutually exclusive} from the
perspective of decision maker, i.e., choosing one alternative means
not choosing any other alternative. Second, the choice set must be
\textit{exhaustive} meaning all alternatives need to be
included. Third, the number of alternatives must be finite. The
last two conditions can be easily met in our case: Our choice set
consists of two packages - {\tt data.table} and {\tt tidy}; Decision
makers are restricted into the group of projects in our collection where
either of those two packages is installed. To ensure the choices
are mutually exclusive we model the choice of the first technology
selected. 

In this paper we  
applied the mixed logit model to study developers' choice over analogous R packages ({\tt data.table} v.s. {\tt tidy}).
While many variations of choice models exist, the mixed logit model
has the fewest assumptions on the distribution of the choice. 
Here we are not trying to solve the classical choice model which, for example, assumes a complete knowledge about the alternatives and produces implicit utility function. Instead, we simply look for factors that strongly affect the decisions developers make, whether these factors may be rational or related to 
cognitive or social biases.
\subsection{Issues}
It's reasonable to believe that the number of issues and how an
issue is solved during the development of a software package may affect a developer's choice. This factor belongs to a set of rational choices.
To measure it we collect the issues reported during the development of {\tt data.table} and {\tt tidy} packages. Since both packages are hosted on GitHub, we use 
GitHub API to scrape all issues\footnote{A pull request is also treated as an issue in this paper. https://developer.github.com/v3/issues/} reported for both packages. We collected 2.6k issues for the {\tt data.table} and 1.6k issues for the {\tt tidy}.
\subsection{Selecting candidates for study of adoption}\label{s:sctc}
We chose software technologies from the data science ecosystem of
projects using the R language because several of the co-authors are
knowledgeable and have decades of development experience in R, and
we, therefore, do not need to seek external experts to provide interpretations of the findings. As with most language-based ecosystems, the core language provides 
only basic functionality with most of the external packages being maintained in CRAN and Bioconductor 
distributions. Each package can be thought as presenting a technology choice. Since the technologies of storing and managing data are crucial in data science, 
we selected two widely used such technologies: {\tt data.table} and {\tt tidy}. 

The extraction of the supply chain data for these two packages started from 1.4M R files in the entire collection, with 70K blobs (versions of these files) 
that contained the statement indicating installation of either package. 
Fewer than 20K commits produced these 70K blobs and were done in 24K (including forks) projects which installed either package.

We further refined the list of projects because a large fraction involved forks of other projects. One of the most typical ways to make contribution to the development of a project on GitHub is by creating a fork for a project, making changes to this clone and then sending a pull request to original project. As a result, a popular project may have hundreds of forked projects that share a large portion of source code and commit history. These forks are not equivalent to the original projects from which these forks were created and we, therefore, need them to 
be removed from consideration. To detect and delete these forks, we classify projects based on common commits, i.e., a pair of projects are linked if they have at least one commit in common. Based on these links a transitive closure 
produces disjoint clusters. After applying such linking method on 24K projects, we obtain 8K clusters. Each cluster represents a single observation 
in our study and, the date when the first blob containing the use of technology was created is used as the date that technology was adopted for this cluster.
\section{Case study}\label{Case study}
Apart from the {\tt dataframe} package that is a part of core R
language, {\tt data.table} and {\tt tidy*} are two other most
popular packages for data manipulation.  
More specifically, {\tt tidy*} represents a list of packages that
share an underlying design philosophy, grammar, and data structures
that are built for data science in R. Hadley Wickham, the Chief
Scientist at RStudio and the main developer of {\tt tidy*},
developed a family of packages called {\tt tidyverse} to facilitate
the usage of {\tt tidy*} packages by assembling them into one meta
package. We extract a set of packages from {\tt tidy*} that share
similar functionalities with {\tt data.table} and refer to all of
them here as the {\tt tidy} package. This includes  {\tt tidyr},
{\tt tibble} and {\tt readr} packages.

{\tt data.table} was written by Matt Dowle in 2008 and is known for
its speed and the ability to handle large data sets. It's an
extension of base R's data.frame with syntax and feature
enhancements for ease of use, convenience and programming
speed. It's built to be a comprehensive, efficient, self-contained
package, to be fast in data manipulation, and it has a succinct
DSL (domain-specific language). Conversely, {\tt tidy} focuses on the beauty of function composition and data layer abstraction which enable users to pull data from different databases using the same syntax.  

To analyze developers' choice of these two packages, we collected and filtered relevant data by leveraging an open source mining infrastructure~\footnote{https://github.com/ssc-oscar/Analytics}~\cite{maworld} which provides not only APIs for extraction of development data on various levels for open source projects, but also intermediate collections/results extremely useful for study of domain knowledge. In particular, as illustrated in Fig.~
\ref{fig} and Sec.~\ref{s:discovery}, we used the collection of all R file names and maps of file-to-blob, blob-to-content, blob-to-commit and commit-to-project to discover our targeted projects. Meanwhile, for each project we found the first commit in which {\tt data.table} or {\tt tidy} was imported by sorting the commit time. By querying the content for this first commit, we came to know the author of this commit, commit message, etc. We set this first commit time as the end point of our analysis for each project, i.e., for every targeted project, we consider explanatory behavior, activities, and relationships as defined to be computed before this end point so that the analysis considers all factors as they were at the time of the analysis.

\subsection{Operationalizing Attributes for Regression Models}\label{ss:measures}
In this section, we operationally define and justify the variables that quantify the key attributes pertaining to the set of software choices available to developers, as well as the characteristics of the developers making the choices.   
\begin{table*}[t]
\fontsize{9}{8}\selectfont
\centering
\caption{Independent variables}
\label{predictors}
\begin{adjustbox}{width=0.9\linewidth}
\begin{tabular}{ l l l l}
\toprule
\textbf{Independent variables} & \textbf{Annotation}                                            & \textbf{Category} & \textbf{Property type}\\ 
\midrule
CumNum                     & the total number of projects that deployed the package              &   exposure                &  choice related                      \\ 
RplGp                       & the time gap until the first reply to an issue &      infectiousness             &  choice related                      \\ 
Unrslvd                     & the number of open issues over the number of all issues        &      infectiousness             &  choice related                       \\ 
StckExch                  & the number of questions with score above 20 related to either package                  &      exposure             &  choice related                       \\ 
C                        & boolean, indicating whether a project contains C file         &      proximity             &  decision maker                       \\ 
Cmts                           & the number of commits                                          &       infectiousness            &  decision maker                      \\ 
Aths                          & the number of authors/developers                               &       infectiousness            &  decision maker                     \\ 
Prx2TD                 & the proximity to tidy through dependency network                &   proximity                &  decision maker                       \\ 
Prx2DT            & the proximity to data.table through dependency network          &   proximity                &  decision maker                        \\ 
AthPrx2TD         & the proximity to tidy through authorship network                &   proximity                &  decision maker                        \\ 
AthPrx2DT    & the proximity to data.table through authorship network          &   proximity                &  decision maker\\
\bottomrule
\end{tabular}
\end{adjustbox}
\end{table*}

We propose 11 variables to measure various aspects that may have influenced a developer's choice. These variables are listed in Table~\ref{predictors}. 

\textbf{Cmts \& Aths} is the number of commits and author captures the size of a project which may affect package adoption. Larger projects, for example, may prefer less
controversial/more conservative choices. This is a quality of the
choice, so it would most closely fit under the ``infectiousness''
category in social contagion theory. We chose not to use lines of code
(LOC) as size measure since it has less stable distribution
than the number of commits, while, at the same time, being highly correlated with it.  Operationally, for a particular adopter, we collected all commits prior to the end point (first adoption of one of 2 targeted packages) by applying project-to-commit map followed by a time point filtering. We extracted the authors of these commits and counted the unique number.

\textbf{CumNum} is the overall number of project deployments ({\tt tidy} or {\tt data.table}) should increase the chance that a developer would be exposed to the usage of a package and may adopt it in their code. This measure falls within the ``exposure'' category of contagion, because it quantifies the chances that a developer may become aware of the technology.
This is characterized as a factor that is not rational, as the project is hypothesized to be biased towards technology that they are more likely to encounter, not technology that would be optimal for that project.  To assess CumNum for each targeted project at the time of potential adoption, we counted the number of adopter projects which had an earlier end point, i.e., the number of projects that already adopted a package ({\tt data.table} or {\tt tidy}) before a decision was made by the developers of package under evaluation. 


\textbf{Unrslvd} issues can indicate package quality. A higher fraction of unresolved issues may indicate that the package has a significant number of problems, which, like a bad review, may undermine people's confidence in it. This is a quality of the choice, so we hypothesize that it relates to the ``infectiousness'' aspect of the social contagion paradigm. To measure this quantity, we leveraged GitHub API to collect all issues for {\tt data.table} and {\tt tidy} packages and filtered issues raised before the decision end point for each adopting project. We count the number of unresolved issues and normalize it over all issues raised before end point, because in general a project tends to have more issues and unresolved ones as its age grows, and we believe the averaged rate of unresolved issues are more reflective of a package's maintenance and quality.

\textbf{C} code being associated with a project is used as a proxy for the requirement for high performance. Typically, computations that are too slow for the interpreted R language are implemented in C to improve performance. This is a requirement of the decision maker that would most closely fit under the ``infectiousness'' category because it likely indicates a strong preference for higher performance embodied by the {\tt data.table} choice. This is a good 
example of a factor that may represent a rational choice for some decision makers. To measure this aspect, we applied commit-to-file map on every commit prior to end point for each project and filtered files with suffix `.[cC]'. 

\textbf{StckExch} is a proxy for the popularity of each package. It counts the number of highly ranked (score $>$ 20) questions related to each package. 
Developers often search for answers to issues they face and may stumble on one of these packages presented as a solution to a problem they are facing, thus increasing chances that they may adopt that technology. From the social contagion perspective this would increase ``exposure''. We avoid counting the total number of questions because most of the questions tend to be of low quality and the search engines tend to avoid including links to them, thus they do not increase ``exposure.''  This factor may be interpreted from a rational perspective 
(leveraging experience of others when lacking other information), but more appropriately, it is 
a great example of social bias since the developer did not engage in due diligence, instead relying on social cues to make a technical choice.  To measure this, we filtered related posts in the StackExchange dump (2017-12-08) by searching {\tt data.table} and {\tt tidy} in post title and content. Furthermore, we filtered posts with high quality by setting post score threshold as 20. Again, we counted posts raised prior to the end point for each adopter project.

\textbf{Prx2DT/Prx2TD} measure dependency networks and can be understood from the perspective of software supply chain networks.  Based on the characteristics of the software domain, especially the open source software community, dependency networks can be viewed as technologies (library/package) spreading from upstream (original package) to downstream (packages where the original package was installed) and, in turn, to further downstream packages. 

We consider all downstream packages of {\tt data.table} and {\tt tidy}, e.g. those in the {\tt data.table} and {\tt tidy} clusters respectively.  We hypothesize that if a project installed a package within the {\tt data.table} cluster, then the project is more likely to install {\tt data.table} than {\tt tidy}. The rationale of such a hypothesis is that if developers installed a package because of 1)preferences for some of its functionalities or features inherited from an upstream package or 2) the way such a package works, which is sometimes influenced by or derived from an upstream package, then it is more likely that these developers will gravitate toward the upstream package over other alternatives. 

Based on the dependencies of R CRAN packages, the clusters of {\tt data.table} and {\tt tidy} are easily constructed. More specifically, we used the METCRAN\footnote{https://www.r-pkg.org/about} API and scraped meta data for more than 11K R CRAN packages for which dependency information is available. Table~\ref{network} summarizes basic information on the networks that were constructed and more detailed information on the methodology follows.


\begin{table}[tbp]
\fontsize{9}{8}\selectfont
\centering
\caption{Network characteristics}
\label{network}
\begin{tabular}{lll}
\toprule
 \textbf{Characteristics}   & \textbf{data.table} & \textbf{tidy} \\ 
\midrule                     
\# downstream packages       & 813                 & 2203          \\ 
\# downstream layers & 5                   & 5             \\ 
\# of packages in common & \multicolumn{2}{c}{636}             \\ 
overlap ratio             & 0.78                & 0.28          \\ 
\bottomrule
\end{tabular}
\end{table}

Each downstream package in the {\tt data.table/tidy} dependency network needs to be weighted before calculating proximity of an adopting package to both {\tt data.table} and {\tt tidy}. We suggest that the algorithm used to determine the weights be based on several key principles:
\begin{itemize}
\item for each downstream package, only the relative weight to the root package ({\tt data.table/tidy}) matters
\item for each downstream package, the sum of its weights to both root packages is a constant
\item the closer to a root package, the higher the weight that a downstream package gets relative to that root package
\end{itemize}
We assume that each package has a weight of 1 in total. Let's denote the packages set in the {\tt data.table} downstream network as $S_d$, that in {\tt tidy}  as $S_t$, the weight of package $a$ to {\tt data.table} as $W_{ad}$ and that to tidy as $W_{at}$, the depth of package $a$ in the {\tt data.table} network as $D_{ad}$ and that in the {\tt tidy} network as $D_{at}$, then based on principles mentioned above, the weights of package $a$ are determined as follows:
\begin{itemize}
\item $W_{ad} = 1, W_{at} = 0 $ if $ a \in S_d$ \& $a \notin S_t$
\item $W_{ad} = 0, W_{at} = 1 $ if $ a \in S_t$ \& $a \notin S_d$
\item otherwise, $W_{ad} = D_{at}/(D_{ad}+D_{at})$, $W_{at} = D_{ad}/(D_{ad}+D_{at})$
\end{itemize}
The next step is to extract the list of packages installed in each observation/project, 
after which we can aggregate the weights of these packages to compute the proximity of each project. 

As we have mentioned in Sec.~\ref{s:ssc}, various maps among Git objects have been created. By utilizing maps of project-To-commit, commit-To-blob, and blob-To-content in sequence and selecting the install statements in blob content via regular expressions similar to those mentioned in Sec.~\ref{s:discovery}, we get the list of packages installed in each project. From this set, we obtain projects that are either in {\tt data.table} or in {\tt tidy} clusters.

For a project $p$, denote the list of packages obtained in last step as $L_p$ and denote a package in that list as $a$. Then the proximity of a project $p$ to {\tt data.table}, denoted as $P_{pd}$, and to {\tt tidy} as $P_{pt}$, can be computed:
\begin{equation}
  \left\{\begin{aligned}
  P_{pd} &= \Sigma_a^{L_p} W_{ad} \\
  P_{pt} &= \Sigma_a^{L_p} W_{at}
\end{aligned}\right.
\end{equation}

To summarize the process described above, we first measured the weight of each downstream package in either {\tt data.table} or {\tt tidy} by leveraging the R package dependency networks and the formulas above. Secondly, by following a similar flow in Fig.~\ref{fig}, we extracted all R packages that were adopted in the commits prior to end point where one of the focal packages was first adopted. Finally, we calculated the proximity to {\tt data.table} and {\tt tidy} by summing up the weights of all downstream packages for each project. Notice that a project's downstream packages that were not in {\tt data.table} or {\tt tidy} downstream set were dropped.

\textbf{AthPrx2DT/AthPrx2TD} represents the proximity of a developer to a focal project as measured through their author network. It can be explained from the perspective of social contagion. Social contagion refers to the propensity for a certain behavior to be copied by others. Consider the fact that developers in GitHub are linked through common projects they are devoted to, where information and ideas are shared and transmitted from one to others, an underlying social network emerges. Organizational actions are deeply influenced by those of other referent entities within a given social system, according to DiMaggio~\cite{10.2307/2095101}: non-adopters are influenced by adopters over time, and they influence the behavior of other non-adopters after their own adoption~\cite{angst2010social} if thinking of our case as package adoption. In short, the adoption of {\tt data.table/tidy} is a temporal process of social contagion.

We attempt to look for developers that are exposed to contagious packages --- {\tt data.table/tidy}. These developers include not only the authors of each package who are directly exposed inherently, but also developers who cooperate with directly-exposed authors in other projects. 
Authors of other projects that are directly exposed to authors of {\tt data.table/tidy}, are identified by applying a project-To-author map to both {\tt data.table/tidy} packages separately and indirectly-exposed authors are obtained by combining the map of author-To-project and the map of project-To-author serially and then applying it on each directly-exposed author. 

We classify authors exposed to {\tt data.table} into the {\tt data.table} author cluster and those exposed to {\tt tidy} into the {\tt tidy} author cluster. Projects/observations may have authors who are in either of these two clusters and these authors may influence the choice of data frame technology, i.e., ({\tt data.table} vs. {\tt tidy}). In order to estimate the impact of every author in each cluster, we use the following weights,
\begin{itemize}
\item $W_{bd} = 1, W_{bt} = 0 $ if $ b \in C_d$ \& $b \notin C_t$
\item $W_{bd} = 0, W_{bt} = 1 $ if $ b \in C_t$ \& $b \notin C_d$
\item otherwise, $W_{bd} = D_{bt}/(D_{bd}+D_{bt})$, $W_{bt} = D_{bd}/(D_{bd}+D_{bt})$
\end{itemize}

where $b$ represents an author in a project; $C_d/C_t$ stands for author cluster of {\tt data.table/tidy}; $D_{bd}/D_{bt}$ refers to the distances from author $b$ to {\tt data.table/tidy}, i.e., author $b$'s depths in the author cluster of {\tt data.table/tidy}, 1 for directly-exposed author and 2 for indirectly-exposed author; $W_{bd}/W_{bt}$ is the proximity of author $b$ to {\tt data.table/tidy}, indicating author $b$'s impact on choosing {\tt data.table/tidy}.
Note that these measures are similar to the ones used in calculating \textit{Prx2DT/Prx2TD} and are based on similar principles. 

After estimating each exposed author's influence, the overall exposed authors' influence in project $p$ can be measured as follows:
\begin{equation}
  \left\{\begin{aligned}
  PA_{pd} &= \frac{\Sigma_b^{A_p} W_{bd}}{N_p} \\
  PA_{pt} &= \frac{\Sigma_b^{A_p} W_{bt}}{N_p}
\end{aligned}\right.
\end{equation}
where $A_p$ is the set of authors of project $p$ who are in either of {\tt data.table/tidy} author cluster; $W_{bd}/W_{bt}$ is the proximity of author $b$ to {\tt data.table/tidy} calculated in previous step; $N_p$ is the number of authors in project $p$;  $PA_{pd}/PA_{pt}$, i.e., \textit{AthPrx2DT/AthPrx2TD}, is the overall influence of exposed authors on a project $p$. Notice that \textit{AthPrx2DT/AthPrx2TD} is calculated through aggregating the influence of each exposed author and being normalized over the total number of authors in that project. The rationale for normalization is that a project tends to have more exposed authors if it contains more authors, resulting in a higher value for \textit{AthPrx2DT/AthPrx2TD}. By normalization we remove this bias induced by the difference in the number of authors for different projects.
This factor falls clearly within a realm of a social bias. It may also be partially explained as cognitive bias if the developer is not aware of alternative choices.

To summarize the computation of proximity through authorship network, we started by measuring the weight of each author who was either a co-author of {\tt data.table/tidy} or had cooperated with at least one of the authors of {\tt data.table/tidy}, which was detailed above. Then we summed up the weight of every author of a project and normalized it over the total number of authors in this project. Again, here we applied end point filter on every step in calculation.

\textbf{RplGp} measures how fast the developers or maintainers of a package respond once an issue has been raised. The timeliness of this response reflects the efficiency of package maintenance and can be attributed to the `infectiousness' category of social contagion theory and could clearly be of interest for those deciding on which package to adopt.

The calculation of reply gap is worth discussing. We are interested in  understanding how long it takes for an issue to get its first reply after being reported.  For each individual in the study, we focus on the time period just before the key commit that includes the choice of focal package ({\tt data.table/tidy}). However, several additional obstacles that needed to be addressed in order to measure the reply gap : 
\begin{enumerate}
\item It is rare that an issue was raised simultaneously with the key commit (inside which either the{\tt data.table/tidy} package is installed).
\item The timeliness of replying to an issue may vary drastically during the development of a package, hence taking the closest issue's reply-time as a representative is not reasonable
\item For some issues, it took a significant amount of time to get a reply and in some cases no reply was ever made to an issue, thus, averaging reply-time to previous issues is problematic due to long right-censored cases.
\end{enumerate}
This is a case where statistical models for survival(time-to-event) are appropriate. In this scenario, an issue can be viewed like a patient under study with the first reply analogous to conclusion of the medical issue or death of the patient. We aim to model the time until reply to the reported issue, i.e., the survival time of the issue, with shorter lifetimes indicating a more interactive development team. Irrespective of package, for each issue, we record the time that it was submitted (timestamp recorded when the issue is raised) and use survival analysis to model the distribution of the issue lifetimes for each package ({\tt data.table/tidy})
using the R package `survival'~\cite{survival}.  Predictions for the reply time for each project (observation) can be made based on data collected before the key commit. The \textit{RplGp} for a project is simply the median issue lifetime for an issue generated before a key commit. This factor appears to be clearly related to rational choice factors as the delays in response may cause real problems.


In practice, we extracted all issues of {\tt data.table/tidy} from GitHub and measured difference between the time an issue is first raised and the first response time. As described above, we trained a survival model to estimate the distribution of the delay until first response delay. The model was trained using all issues that had been raised before the current package key commit. Those issues that had not been responded to yet were right censored in the model fitting.  The reply gap represents the median value of response times.

In summary, we note that for each project that eventually adopts one of the two focal packages ({\tt data.table/tidy}), all of the variables described in this section are calculated dynamically using only data that occurs before the key commit.  In addition, for each observation, every predictor with choice property (Table~\ref{predictors}) needs to be calculated for both packages, e.g., \textit{Unrslvd} needs to be calculated for both {\tt data.table} and {\tt tidy}. These will end up being denoted as \textit{Unrslvd.datatable} and \textit{Unrslvd.tidy}.

\section{Results}\label{s:results}
We identified 24,000 projects (7,000 for {\tt tidy}, 17,000 for {\tt data.table}) that installed either {\tt data.table} or {\tt tidy} between June, 2009 and January, 2018. After removing forks, we were left with a total of 8,000 projects (3,000 for {\tt tidy}, 5,000 for {\tt data.table}). Furthermore, we removed project adoptions occurring prior to June 16, 2014, when {\tt tidy} was first introduced. Before then and only {\tt data.table} existed as a viable option, so no choice was possible. As a result, we dropped approximately 20\% of 5,000 data.table observations representing adoptions prior to June 16, 2014. 
The remaining projects serve as observations in our choice model. 
Table~\ref{Input statics} summarizes basic statistics for independent variables analyzed in the model. 
We use the R package 'mlogit'\footnote{\url{https://cran.r-project.org/web/packages/mlogit/vignettes/mlogit.pdf}}~\cite{mlogit} to fit the model using the 11 predictor variables defined above with the response being an indicator of the package chosen. 

Very high correlations among predictors (above 0.9) occurred
between {\em Prx2DT} and {\em Prx2TD}. High
correlations may lead to unstable and difficult to interpret models and
need to be addressed.
Since we do not have any {\em a priori} theory-derived reasoning for
removing one or the other variable, we removed {\em Prx2DT}. The modeling results remain stable if this approach is reversed.
Table~\ref{result:noncumulative} presents the resulting model fit.
\begin{table}[ht]
\fontsize{9}{8}\selectfont
\centering
\caption{Summary Statistics for Independent variables}
\label{Input statics}
\begin{tabular}{l l l l l}
\toprule
 \textbf{Variable}          & \textbf{median}       & \textbf{mean}         & \textbf{std.dev}     \\ 
\midrule
Cmts                        & 3  & 46.83  & 645.68  \\           
 Aths                         &     1   &  2.13   & 8.23  \\ 
 C (boolean)                      &        0   & 9.79e-03  & 9.85e-02 \\
 Prx2DT         &        0    &   0.15   &   0.95    \\ 
 Prx2TD              &        0     &    0.62  &  2.79     \\ 
 AthPrx2DT  &      0     & 6.99e-2   & 0.17   \\ 
 AthPrx2TD      &     0     &  0.11  &  0.24  \\ 
 CumNum.datatable         &     2.72e+03    & 2.66e+03   & 1.87e+03   \\ 
 CumNum.tidy           &     305    & 8.44e+02   & 9.12e+02  \\ 
 RplGp.datatable          &     1.41     & 1.42    & 0.21  \\ 
 RplGp.tidy               &     1.95     & 1.98    & 0.31     \\ 
 Unrslvd.datatable        &     0.27    & 0.28   & 4.00e-2  \\ 
 Unrslvd.tidy           &     0.15     & 0.19    & 7.46e-2  \\ 
 StckExch.datatable    &     130    & 125.76   & 6.53 \\ 
 StchExch.tidy         &     158    & 152.57   & 10.14 \\ 
 \bottomrule
\end{tabular}
\end{table}
\begin{table}[htp]
\centering
\fontsize{9}{8}\selectfont
\caption{The Fitted Coefficients.\\ $McFadden$\cite{mcfadden1973conditional} $R^2=0.14$ $n=7k$}
\label{result:noncumulative}
\begin{tabular}{l l l l}
\toprule
\textbf{Variable} & \textbf{Estimate}    &  \textbf{Std. Error}  &    \textbf{p-val}  \\
\midrule
tidy:(intercept) & -6.72  & 0.26            & 2.20e-16\\ 
CumNum           & 1.43e-04  & 1.43e-05    & 2.20e-16 \\ 
Unrslvd       & 8.30e-02 & 0.72            & 0.91 \\ 
RplGp         & -0.33 & 7.94e-02            & 2.52e-08 \\ 
StckExch    & 0.25 & 0.01             & 2.20e-16\\ 
tidy:Cmts        & -4.63e-04 & 2.26e-04             &  5.64e-2 \\ 
tidy:Aths         & 8.53e-05  & 7.17e-03             &  0.99   \\ 
tidy:C       & -0.64  & 0.28            & 2.40e-3  \\ 
tidy:Prx2TD  & 0.18 & 2.89e-02           & 6.78e-10   \\
tidy:AthPrx2TD  & 1.27 & 0.14    &  2.20e-16        \\ 
tidy:AthPrx2DT      & -9.03e-02 & 0.19 & 0.65   \\ 
\bottomrule
\end{tabular}
\end{table}

Below we summarize findings for each predictor variable separately.

\textbf{StckExch}: the coefficient is 0.2, indicates that the number of high quality 
questions on StackExchange is associated with the 
likelihood that a project would adopt the respective technology.  The association is positive, holding
other factors equal.  For illustration, if the number of high quality questions increases by 6 questions (1 std. dev.) from a median value of 130 for {\tt data.table}, the estimated probability of choosing
{\tt data.table} increases from 0.58 to 0.87, while holding all other predictors at their median values.

This result aligns well with the social contagion theory that posits that
increased adoption is a consequence of increased exposure.  Surprisingly, including an additional predictor that counts the total number of questions (of high and low quality), shows no statistical significance. It appears to be counter-intuitive as more exposure should increase adoption. 
However,  when developers want to solve an issue related to the functionality of the R {\tt data.frame}, they often may not search on StackExchange, but use a general search engine and follow links to StackExchange. The total number of posts, therefore,  may be not visible to developers, only the set of posts that the search engine deems to be of sufficiently high quality. The number of posts (questions), may, therefore, not be a good proxy of exposure. As such, the total number of posts of Low-quality questions, in fact, appear to discourage developers from using a package. 
\hypobox{\N{\textbf{Finding 1:} \textit{We found that exposure measured via the total number of questions on StackExchange had no impact on adoption, while the number of high quality questions has a strong and positive correlation with increased adoption.}}}

\hypobox{\N{\textbf{Finding 2:} \textit{We did not find statistically significant association between infectiousness as measured via the fraction of unresolved issues and adoption rates}}}

\textbf{Unrslvd}: the ratio of unresolved issues over total number of issues does not show a statistically significant association with the choice between the two technologies. This might be due to the fact that the ratio of unresolved issues can not be easily observed by adopters, e.g. the default number in the issues tab on GitHub represents the current number of issues that are not resolved and adopters have to take a further step to calculate such ratio. Another explanation is that adopters may investigate influential unresolved issues instead of all unresolved issues, which suggests that more detailed issues analysis such as segmentation and classification are required to understand adopters' decision making process more precisely.


\textbf{AthPrx2TD}: the coefficient is 1.3, indicating that the closer a project is to authors of the package {\tt tidy} vis-a-vis the author network, the more likely they are to choose {\tt tidy} over {\tt data.table}. If the proximity to {\tt tidy} in the author network increases by one standard deviation of 0.24 from a median value of 0 (e.g., a project that has four authors and one of them cooperates with {\tt tidy}'s developers, but not with any of {\tt data.table}'s developers), the estimated probability of choosing {\tt tidy} increases seven percent from 0.42 to 0.49. This finding supports the basic premise of the social contagion hypothesis that developers' choices are affected by the environment they are in. 
\hypobox{\N{\textbf{Finding 3:} \textit{Proximity as measured by the fraction of authors who are either developers of the package to be adopted or who work with at least one developer of that package, increase the chances of adoption.}}}
 This may be a consequence of authors who have direct expeience or are familiar through word-of-mouth. However, \textbf{AthPrx2DT} is not statistically significant. One reason may be that {\tt data.table} is a more widely deployed package and the deployments may play 
a larger role than the social connections. Also, each community of users and developers may be different. For example, the {\tt tidy} community may have more social interactions
than the {\tt data.table} community. Furthermore, the exposure in the {\tt tidy} community may come from a much larger set of packages in the {\tt tidyverse}, while {\tt data.table} does not have an equivalent brand that involves a wider variety of tools beyond data handling. 

\textbf{C}: the coefficient is -0.6, indicating that a project containing at least one C file is less likely to choose {\tt tidy}. The estimated chances of choosing {\tt data.table} increase by 15 percent from 0.58 to 0.73. The finding is consistent with our hypothesis that if an R project has a 
need for performance, as evidenced by the use of functionality being developed natively in C language, then it is more likely to to choose the higher performance of {\tt data.table}. 
\hypobox{\N{\textbf{Finding 4:} \textit{Proximity, as measured by the project's need for performance, is associated with adoption of packages that emphasize high performance}}}

\textbf{RplGp}: the coefficient is around -0.4, indicating that the more quickly a package's issue gets a response, the more likely that this package will be chosen. If the number of days until first response to an issue increases by 0.21 days (1 std. dev.)  from a median value of 1.4 for {\tt data.table}, the estimated chances of a project choosing {\tt data.table} decrease by two percent from 0.58 to 0.56 assuming all other variables remain at their median values.
The time until first response is not as readily visible to developers as most other measures that we used, so developers may not be able to observe it when making a choice. 
However, it appears to be a reasonable 
proxy for project's reactions to external requests that could be easily gleaned by reading through some of issues on the issue tracker. 
A well maintained package is more likely to respond to new issues quickly and thoroughly, leaving a good impression and, thus, increasing the likelihood of being adopted. 
This has implications for designing project dashboards intended to make key project attributes more visible. 
\hypobox{\N{\textbf{Finding 5:} \textit{Infectiousness of a package as measured by speed of response to issues is associated with a higher adoption rate for that package.}}}

\textbf{Prx2TD}: the coefficient is 0.2, indicating that the closer (through a dependency network) a project is to the package {\tt tidy}, the more likely its authors are to choose {\tt tidy} over {\tt data.table}. If proximity to {\tt tidy} in dependency network increases by one standard deviation of 2.8  from a median value of 0 (e.g., a project installs/uses three packages that are in first layer downstream from {\tt tidy}), the chances of choosing {\tt tidy} go up by 12 percent from 0.42 to 0.54. It supports our hypothesis that the supply chain influences projects' choices. 
A project tends to install a specific package if it has already installed other packages that also depend on it, i.e., if a project uses downstream dependencies of a package, it is more likely to use the package itself rather than other alternatives. Being familiar with downstream packages may reduce the overhead or learning curve required for an upstream package, leading to an advantage over other choices.
\hypobox{\N{\textbf{Finding 6:} \textit{Proximity to a package as measured via technical dependency networks is associated with a higher adoption rate.}}}

\textbf{CumNum}: the coefficient is 1.4e-4, indicating that a larger number of deployments of a package in the past will make it more likely to be adopted. If the number of deployments increases by one standard deviation, 1870
projects, from a median value of 2660 projects for {\tt data.table}, the estimated chances of choosing
{\tt data.table} go up by seven percent from 0.58 to 0.65 for a project holding all other values at the median.
A larger number of overall deployments, on one hand, increases the chance for a package to be known by adopters. On the other hand, from the perspective of adopters, more deployments usually insinuate a stable and mature product (though it is not clear if the number of deployments is visible to a developer), and enhances adopters' confidence in this package. Either of these reasons justifies adoption of the widely deployed package as predicted by the social contagion theory.

\hypobox{\N{\textbf{Finding 7:} \textit{Exposure to a package that is widely deployed is associated with a higher adoption rate.}}}


\hypobox{\N{\textbf{Finding 8}: We did not find statistically significant association between infectiousness as measured via the number of commits and adoption propensity \footnote{the coefficient is not statistically significant at 0.005 level recommended 
for reproducibility; see Table 4 and ~\cite{benjamin2018redefine}.}}}

We also find that the number of authors in the adopting project does not affect the choice of technologies. Social contagion theory does not suggest that this predictor should have an effect, but it could be that project activity (which has a substantial correlation with the number of authors), may already account for the differences in propensity to chose {\tt tidy} over {\tt data.table} making the variation in the number of authors statistically insignificant. 
\vspace{-.025in}
We achieved a $McFadden$\footnote{https://stats.stackexchange.com/questions/82105/mcfaddens-pseudo-r2-interpretation}  $R^2$ of 0.14, which is a good fit according to McFadden\cite{mcfadden1973conditional}. (Notice that the R package `mlogit' use $McFadden$ $R^2$ instead of $R^2$ to estimate fitness of the model because logit models don't generate the sums-of-squares needed for standard $R^2$ calculation.)

Regression models are explanatory, but we can also use them to do
prediction. The 10-fold cross-validation done by randomly splitting projects
into 10 parts and fitting the model with predictors listed
in Table~\ref{result:noncumulative} on nine parts and predicting on the
remaining part yielded a reasonable AUC of 73\%. Average accuracy
was 70\% with balanced Type I and II errors (obtained by choosing
predicted probability cutoff of 0.49).

Finally, it is worth noting that out of six predictors that were statistically significant, 
only CumNum, RplGp, and C were clearly grouped into predictors that would support 
rational choice. The remaining three predictors primarily reflect a mixture of social and cognitive 
biases associated with social preference or default choice when alternatives are not known. 
If we include the effort needed to obtain the necessary information into the utility function, these 
social and cognitive biases can, of course, be explained rationally as well.

In a parallel study to the regression analysis discussed above, surveys were emailed to 1085 of the developers that
had committed R projects containing the {\tt data.table} and {\tt tidy} libraries.  The purpose of the 12-question survey was to gain insights into the reasons behind the user's selection of a particular package. 
One of the survey questions asked the 803 respondents how important 13 common factors were
when selecting a package to expand the function of the basic R data.frame. Respondents
selected a factor-tile, such as ``Package's Historic Reputation'', and moved it to the box
that represented the priority for that user.  A user could select/drag as
many or as few tiles as they wanted. 

\begin{figure}[h!]
\centerline{\includegraphics[width=10cm,height=8cm,keepaspectratio]{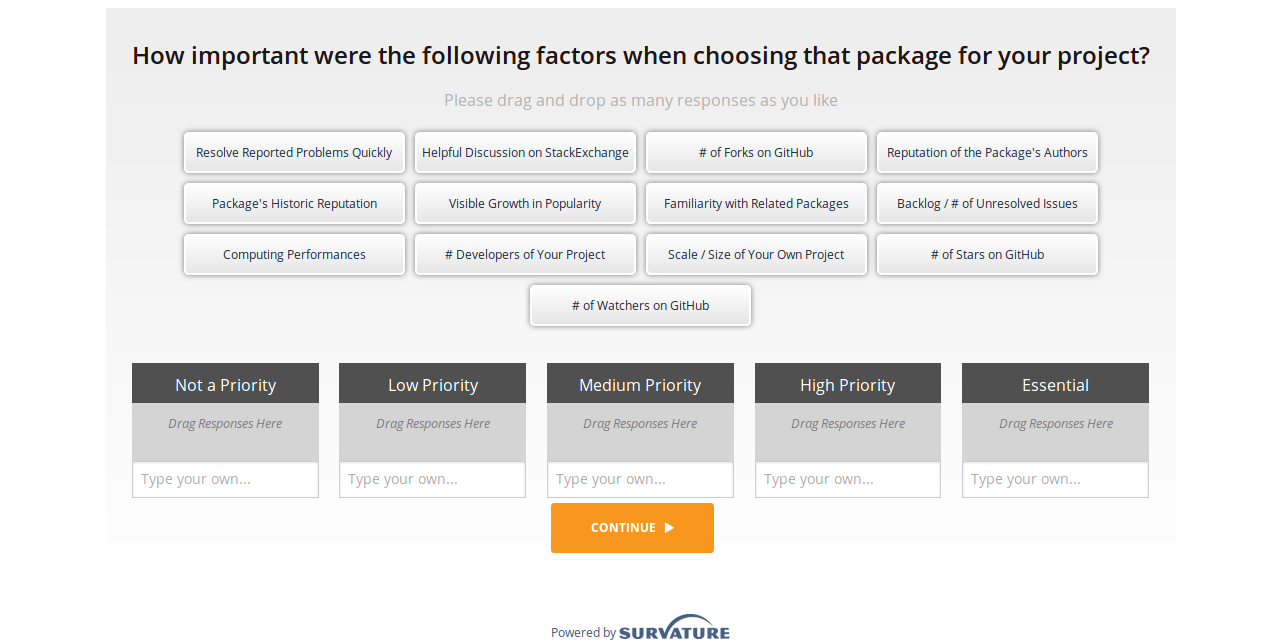}}
\caption{Survey Question 5}
\end{figure}

In agreement with the presented regression, both {\tt data.table} and {\tt tidy} users indicated that they value Stack Exchange as a technical reference. {\tt tidy} users place importance on the reputation of the author and the package, reinforcing the sentiment associated with the significant regression variables Prx2TD and AthPrx2TD. The {\tt data.table} users are concerned with computing performance, bolstering the regression findings that a project with at least one C file is likely to choose {\tt data.table}. (Note - The
shortened labels below correspond to the order of the tiles above.  The full summary of the survey results can be found by following the hyperlink listed after Section 8) 

\begin{figure}[h!]
\centerline{\includegraphics[width=10cm,height=8cm,keepaspectratio]{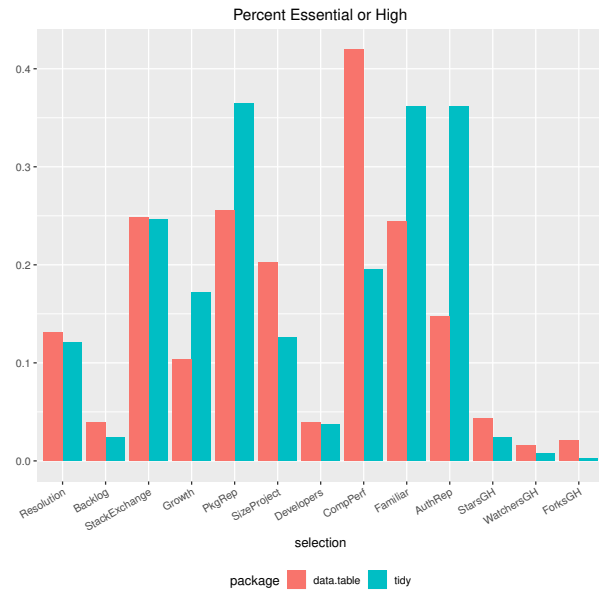}}
\caption{Survey Question 5 Summary}
\end{figure}
\section{Limitations}\label{s:limitations}
Empirical studies must be interpreted carefully due to a number
of inherent limitations. Here we highlight some of the potential
issues and how we tried to address them. 

To obtain an unbiased picture of technology spread representing the
entirety of the projects, we consider a very large collection of
projects. While large, our sample cannot be complete as many
projects do not publish their code and our collection may have
missed even some of the public projects.  The
sample we have should limit the findings to projects that share
their version control data on one of the many forges, such as
GitHub, BitBucket, GitLab, Bioconductor, SourceForge, etc. However,
it may not be representative of the entire universe of projects,
especially projects that do not publish their version control data.

We have selected only projects with extension $[rR]$, but some older
projects may use extension $[sS]$ indicating the historic name for  
R language, or some other source code without any (known) extension.  

Regular expressions that we used to identify instances of package usage or installation can capture most of the install statement in the .r/R file, however, in some cases the install statement may be missed due to a dynamic specification in the installation such as in the case below, 
\begin{lstlisting}
ipak <- function(pkg){
    new.pkg <- pkg[!(pkg %in% installed.packages()[, "Package"])]
    if (length(new.pkg)) install.packages(new.pkg, dependencies = TRUE)
    sapply(pkg, require, character.only = TRUE)
}
# usage
packages <- c("ggplot2", "plyr", "reshape2", "RColorBrewer", "scales", "grid")
ipak(packages)
\end{lstlisting}
Also, multiple packages may be wrapped into a variable before calling the install function:
\begin{lstlisting}
load.lib<-c("EIAdata", "gdata", ...,"stringr","XLConnect", 
"xlsReadWrite","zipcode")
install.packages(lib,dependences=TRUE)
\end{lstlisting}

Moreover, regular expressions may falsely capture an install statement in
some cases, e.g., install statements that are commented out may, in rare cases, be captured by regular expressions. Files that are contained in a project but not used may also contain installment statements that are captured by regular expressions. To alleviate this issue we used the R language requirement to have a comment character '\#'  on each line and ensured that the matched install is never preceded by the comment character.

These potential errors would affect the dependency networks we construct and
result in an under/over-count of the number of projects using the packages we
are looking for.  Developer identities may not be spelled the same
way and that may affect the author
network~\cite{Bird:2006:MES:1137983.1138016}.  We have tried to address
these and other issues encountered when dealing with operational
data from software repositories and big
data~\cite{Mockus:2014:EBD:2593882.2593889,ms2007software,GBM16}.

In order to validate our findings in domains other than R, we applied our model and approach on JavaScript frameworks (Angular, React) and achieved similar results\footnote{https://github.com/ssc-oscar/PackageAdoptionAnslysis}. We expect that more cases will be studied in the future testing external generalizability of this approach. 

Its important to note that the particular operationalizations of the
concepts from social contagion theory are just a few of the possibilities. Measures are not entirely orthogonal: each measure may capture the aspects of other dimensions
beyond the one it is intended to measure. The correlations among
predictors may lead to unstable models that are hard to
interpret. We address this limitation by carefully considering various
interpretations of the measures, conducting exploratory analysis of
the obtained measures, selecting a subset that does not pose threats
to model stability and investigating compliance with model
assumptions including inspection of outliers, non-homogeneous
variance, and performing general model diagnostics. We also
model the first choice, but it is also reasonable to model
the full set of choices made. In the latter case, we need to include
the third option, i.e., projects choosing both packages: {\tt tidy} and
{\tt data.table}. 
We fitted a variety of alternatives to ensure that the reported
results are not affected by these variations in the approach.
 We only present the results for two alternatives due to space
considerations, but we have applied choice model on several other R
packages as well.

Finally, here we demonstrate how to use the social contagion modeling 
using version control data to evaluate developer behaviour when choosing software packages. 
The particular results we obtained for R and the specific two packages may not, therefore generalize
beyond this specific context. Only additional studies in multiple contexts can validate the generality of these 
findings.

\section{Related Work}\label{s:related}

The closest related work involves studies of use and migration of software
libraries. A number of metrics and approaches were proposed to
 mine and explore usage and migration trends.
A software library encapsulates certain functionality that is then
used by applications (or other libraries). The application may
benefit from extra functionality or performance in the new libraries
that may be created later, but switching to a new library (library migration) involves
some recoding of the application\cite{hora2015apiwave,mileva2010mining,Cossette:2012:SGT:2393596.2393661,Lammel:2011:LAA:1982185.1982471,Nguyen:2010:GAA:1932682.1869486}.  
Most prior work, therefore, focused on costs and benefits of
library migration~\cite{kabinna2016logging,teyton2014study,de2018library,teyton2012mining,Tonelli2010,Tonelli2011,Dagenais2009,Mileva:2009:MTL:1595808.1595821}. 
Similarly to that work we ask why developers chose a new library. In contrast 
to prior work, we construct new predictors of adoption (e.g.,  technical and author dependency
networks, breadth of deployment, quality of support measured
through StackExchange, issue number and response times) 
that are based on sound theoretical foundations and we use choice models to understand how macro trends
at the scale of the entire SSC emerge from actual decisions the
individual developers make to select a specific software technology.

Approaches to detect library usage include issue report
analysis~\cite{kabinna2016logging}. As in prior work we detect usage by searching for
library statements in source files of
projects~\cite{teyton2014study}.  
 De la Mora \textit{et al.}~\cite{de2018library} introduce
an interface to help developers choose among the
libraries by displaying their popularity, release
frequency, and recency.  While building on this research, we add novel network, deployment, and
quality measures that would inform developer choice. More importantly, we 
radically improve the ability
of developers' to make informed decisions by providing a statistical model that explains which of these measures matter and how they affect the choice.

Prior studies that examined technology choices have used a variety of approaches ranging from surveying developer preferences~\cite{anderson1990choice} and reasons~\cite{xiao2014social} behind, to mining version control and issue tracking repositories~\cite{kabinna2016logging, teyton2014study, de2018library}. Similarly, we mine version control data, but at a larger scale of all projects with public version control data that include R language files. This allows us to construct complete software supply chains that depict end-to-end technical and social dependencies.

\section{Conclusions}\label{s:conclusions}

Integrating software supply chain concepts and models to operationalize key variables from social contagion theory to investigate software technology adoption appears to have provided a number of potentially useful insights in the present case study of two data manipulation technologies within R language. 
More specifically, the methodology was able to identify factors that were influential in decision-makers' choices between software technologies and demonstrate the need to account, not only for the  properties of the choice, but also of the chooser and of the importance of the supply chain dependencies and information flows. It also validates the measures deemed to be the drivers of technology adoption by the social contagion theory. 

This study introduces the concept of two types of software supply chains (based on technical dependencies and on the relationships among developers induced by projects they have worked on) and demonstrates how software supply chains for 
the entire open source ecosystem can be reconstructed as they have existed at any point in the past from public version control systems. Additionally, by taking a social contagion perspective and employing the logistic regression models, we explicate a parsimonious model that is capable of modeling software technology choices. The findings of this study have wide reaching implications for the software engineering community as well as those who study traditional supply chains. For example, the ability to model and understand which aspects of a network of software supply chain or physical supply chain partners and affiliates influence uptake and spread of a given artifact (e.g., package or product) might help contributors adjust their contributions in a way to maximize their reach, while also extending the viability and propagation of a core package or product. This notion is consistent with our findings that a number of characteristics of a developer and properties of technology are found to be important in the choice between major alternatives. More specifically, packages with large number of overall adopters, higher responsiveness to new issues, and more high-quality stack exchange questions are more likely to be chosen. 
Furthermore, from the perspective of a project's decision-makers, their technical features and proximity to a technology in both the technical dependency network and author collaboration network increase the probability of adoption. On a more speculative side, we find that half of the significant predictors do not appear to be related to a traditional 
rational choice, but are likely a reflection of social and cognitive biases or, in plain language, 
shortcuts people take. Developers, at least in the context of technical decisions regarding which package to use, do not appear to be immune from these biases.

Source code and data for this study is publicly available \footnote{\url{https://drive.google.com/drive/folders/1YjC31l5NrD5XzI5ZyxtRLF29OM2owb1X?usp=sharing}} to facilitate reproducibility and 
wider adoption of the proposed methodology.

\section*{Acknowledgment}
This work was supported by the National Science Foundation NSF Award  IIS-1633437.





\newpage
\ifCLASSOPTIONcaptionsoff
  \newpage
\fi



\bibliographystyle{IEEEtran}
\bibliography{references}

\begin{thebibliography}{10}
\providecommand{\url}[1]{#1}
\csname url@samestyle\endcsname
\providecommand{\newblock}{\relax}
\providecommand{\bibinfo}[2]{#2}
\providecommand{\BIBentrySTDinterwordspacing}{\spaceskip=0pt\relax}
\providecommand{\BIBentryALTinterwordstretchfactor}{4}
\providecommand{\BIBentryALTinterwordspacing}{\spaceskip=\fontdimen2\font plus
\BIBentryALTinterwordstretchfactor\fontdimen3\font minus
  \fontdimen4\font\relax}
\providecommand{\BIBforeignlanguage}[2]{{%
\expandafter\ifx\csname l@#1\endcsname\relax
\typeout{** WARNING: IEEEtran.bst: No hyphenation pattern has been}%
\typeout{** loaded for the language `#1'. Using the pattern for}%
\typeout{** the default language instead.}%
\else
\language=\csname l@#1\endcsname
\fi
#2}}
\providecommand{\BIBdecl}{\relax}
\BIBdecl

\bibitem{von2001innovation}
E.~Von~Hippel, ``Innovation by user communities: Learning from open-source
  software,'' \emph{MIT Sloan management review}, vol.~42, no.~4, p.~82, 2001.

\bibitem{VONKROGH20031217}
\BIBentryALTinterwordspacing
G.~von Krogh, S.~Spaeth, and K.~R. Lakhani, ``Community, joining, and
  specialization in open source software innovation: a case study,''
  \emph{Research Policy}, vol.~32, no.~7, pp. 1217 -- 1241, 2003, open Source
  Software Development. [Online]. Available:
  \url{http://www.sciencedirect.com/science/article/pii/S0048733303000507}
\BIBentrySTDinterwordspacing

\bibitem{doi:10.1093/oxrep/17.2.248}
\BIBentryALTinterwordspacing
B.~Kogut and A.~Metiu, ``Open‐source software development and distributed
  innovation,'' \emph{Oxford Review of Economic Policy}, vol.~17, no.~2, pp.
  248--264, 2001. [Online]. Available:
  \url{http://dx.doi.org/10.1093/oxrep/17.2.248}
\BIBentrySTDinterwordspacing

\bibitem{west2006patterns}
J.~West and S.~Gallagher, ``Patterns of open innovation in open source
  software,'' \emph{Open Innovation: researching a new paradigm}, vol. 235,
  no.~11, 2006.

\bibitem{holdsworth1995software}
J.~Holdsworth, \emph{Software Process Design}.\hskip 1em plus 0.5em minus
  0.4em\relax McGraw-Hill, Inc., 1995.

\bibitem{farbey1999exploiting}
B.~Farbey and A.~Finkelstein, ``Exploiting software supply chain business
  architecture: a research agenda,'' 1999.

\bibitem{huang2002product}
\BIBentryALTinterwordspacing
S.~H. Huang, M.~Uppal, and J.~Shi, ``A product driven approach to manufacturing
  supply chain selection,'' \emph{Supply Chain Management: An International
  Journal}, vol.~7, no.~4, pp. 189--199, 2002. [Online]. Available:
  \url{https://doi.org/10.1108/13598540210438935}
\BIBentrySTDinterwordspacing

\bibitem{kalish1985new}
\BIBentryALTinterwordspacing
S.~Kalish, ``A new product adoption model with price, advertising, and
  uncertainty,'' \emph{Management Science}, vol.~31, no.~12, pp. 1569--1585,
  1985. [Online]. Available: \url{http://www.jstor.org/stable/2631795}
\BIBentrySTDinterwordspacing

\bibitem{russell2004people}
D.~M. Russell and A.~M. Hoag, ``People and information technology in the supply
  chain: Social and organizational influences on adoption,''
  \emph{International Journal of Physical Distribution \& Logistics
  Management}, vol.~34, no.~2, pp. 102--122, 2004.

\bibitem{christopher2004mitigating}
\BIBentryALTinterwordspacing
M.~Christopher and H.~Lee, ``Mitigating supply chain risk through improved
  confidence,'' \emph{International Journal of Physical Distribution \&
  Logistics Management}, vol.~34, no.~5, pp. 388--396, 2004. [Online].
  Available: \url{https://doi.org/10.1108/09600030410545436}
\BIBentrySTDinterwordspacing

\bibitem{angst2010social}
C.~M. Angst, R.~Agarwal, V.~Sambamurthy, and K.~Kelley, ``Social contagion and
  information technology diffusion: the adoption of electronic medical records
  in us hospitals,'' \emph{Management Science}, vol.~56, no.~8, pp. 1219--1241,
  2010.

\bibitem{SAMADI2016263}
\BIBentryALTinterwordspacing
M.~Samadi, A.~Nikolaev, and R.~Nagi, ``A subjective evidence model for
  influence maximization in social networks,'' \emph{Omega}, vol.~59, pp. 263
  -- 278, 2016. [Online]. Available:
  \url{http://www.sciencedirect.com/science/article/pii/S0305048315001425}
\BIBentrySTDinterwordspacing

\bibitem{chhajed2005software}
A.~A. Chhajed and S.~H. Xu, ``Software focused supply chains: Challenges and
  issues,'' in \emph{Industrial Informatics, 2005. INDIN'05. 2005 3rd IEEE
  International Conference on}.\hskip 1em plus 0.5em minus 0.4em\relax IEEE,
  2005, pp. 172--175.

\bibitem{ellison2010supply}
R.~J. Ellison and C.~Woody, ``Supply-chain risk management: Incorporating
  security into software development,'' in \emph{System Sciences (HICSS), 2010
  43rd Hawaii International Conference on}.\hskip 1em plus 0.5em minus
  0.4em\relax IEEE, 2010, pp. 1--10.

\bibitem{bass1969new}
\BIBentryALTinterwordspacing
F.~M. Bass, ``A new product growth for model consumer durables,'' \emph{Manage.
  Sci.}, vol.~50, no. 12 Supplement, pp. 1825--1832, Dec. 2004. [Online].
  Available: \url{http://dx.doi.org/10.1287/mnsc.1040.0264}
\BIBentrySTDinterwordspacing

\bibitem{rogers1995innovation}
E.~M. Rogers, ``Innovation in organizations,'' \emph{Diffusion of innovations},
  vol.~4, pp. 371--404, 1995.

\bibitem{fichman2004going}
R.~G. Fichman, ``Going beyond the dominant paradigm for information technology
  innovation research: Emerging concepts and methods,'' \emph{Journal of the
  association for information systems}, vol.~5, no.~8, p.~11, 2004.

\bibitem{10.2307/2095101}
\BIBentryALTinterwordspacing
P.~J. DiMaggio and W.~W. Powell, ``The iron cage revisited: Institutional
  isomorphism and collective rationality in organizational fields,''
  \emph{American Sociological Review}, vol.~48, no.~2, pp. 147--160, 1983.
  [Online]. Available: \url{http://www.jstor.org/stable/2095101}
\BIBentrySTDinterwordspacing

\bibitem{Burt:social}
\BIBentryALTinterwordspacing
R.~S. Burt, ``Social contagion and innovation: Cohesion versus structural
  equivalence,'' \emph{American Journal of Sociology}, vol.~92, no.~6, pp.
  1287--1335, 1987. [Online]. Available: \url{https://doi.org/10.1086/228667}
\BIBentrySTDinterwordspacing

\bibitem{dabbish2012social}
\BIBentryALTinterwordspacing
L.~Dabbish, C.~Stuart, J.~Tsay, and J.~Herbsleb, ``Social coding in github:
  Transparency and collaboration in an open software repository,'' in
  \emph{Proceedings of the ACM 2012 Conference on Computer Supported
  Cooperative Work}, ser. CSCW '12.\hskip 1em plus 0.5em minus 0.4em\relax New
  York, NY, USA: ACM, 2012, pp. 1277--1286. [Online]. Available:
  \url{http://doi.acm.org/10.1145/2145204.2145396}
\BIBentrySTDinterwordspacing

\bibitem{tsay2014influence}
\BIBentryALTinterwordspacing
J.~Tsay, L.~Dabbish, and J.~Herbsleb, ``Influence of social and technical
  factors for evaluating contribution in github,'' in \emph{Proceedings of the
  36th International Conference on Software Engineering}, ser. ICSE 2014.\hskip
  1em plus 0.5em minus 0.4em\relax New York, NY, USA: ACM, 2014, pp. 356--366.
  [Online]. Available: \url{http://doi.acm.org/10.1145/2568225.2568315}
\BIBentrySTDinterwordspacing

\bibitem{mcfadden1973conditional}
D.~McFadden \emph{et~al.}, ``Conditional logit analysis of qualitative choice
  behavior,'' 1973.

\bibitem{choicemodelKa}
\BIBentryALTinterwordspacing
W.~A. Kamakura and G.~J. Russell, ``A probabilistic choice model for market
  segmentation and elasticity structure,'' \emph{Journal of Marketing
  Research}, vol.~26, no.~4, pp. 379--390, 1989. [Online]. Available:
  \url{http://www.jstor.org/stable/3172759}
\BIBentrySTDinterwordspacing

\bibitem{HAUSMAN19951}
\BIBentryALTinterwordspacing
J.~A. Hausman, G.~K. Leonard, and D.~McFadden, ``A utility-consistent, combined
  discrete choice and count data model assessing recreational use losses due to
  natural resource damage,'' \emph{Journal of Public Economics}, vol.~56,
  no.~1, pp. 1 -- 30, 1995. [Online]. Available:
  \url{http://www.sciencedirect.com/science/article/pii/0047272793014157}
\BIBentrySTDinterwordspacing

\bibitem{Talluri:2004:RMU:989113.989136}
\BIBentryALTinterwordspacing
K.~Talluri and G.~van Ryzin, ``Revenue management under a general discrete
  choice model of consumer behavior,'' \emph{Manage. Sci.}, vol.~50, no.~1, pp.
  15--33, Jan. 2004. [Online]. Available:
  \url{http://dx.doi.org/10.1287/mnsc.1030.0147}
\BIBentrySTDinterwordspacing

\bibitem{Tim2004}
\BIBentryALTinterwordspacing
T.~J. Gilbride and G.~M. Allenby, ``A choice model with conjunctive,
  disjunctive, and compensatory screening rules,'' \emph{Marketing Science},
  vol.~23, no.~3, pp. 391--406, 2004. [Online]. Available:
  \url{https://doi.org/10.1287/mksc.1030.0032}
\BIBentrySTDinterwordspacing

\bibitem{mcfadden2000mixed}
D.~McFadden and K.~Train, ``Mixed mnl models for discrete response,''
  \emph{Journal of Applied Econometrics}, vol.~15, no.~5, pp. 447--470.

\bibitem{berry1994estimating}
\BIBentryALTinterwordspacing
S.~T. Berry, ``Estimating discrete-choice models of product differentiation,''
  \emph{The RAND Journal of Economics}, vol.~25, no.~2, pp. 242--262, 1994.
  [Online]. Available: \url{http://www.jstor.org/stable/2555829}
\BIBentrySTDinterwordspacing

\bibitem{Kenneth1981}
\BIBentryALTinterwordspacing
K.~Small and H.~Rosen, ``Applied welfare economics with discrete choice
  models,'' \emph{Econometrica}, vol.~49, no.~1, pp. 105--30, 1981. [Online].
  Available:
  \url{https://EconPapers.repec.org/RePEc:ecm:emetrp:v:49:y:1981:i:1:p:105-30}
\BIBentrySTDinterwordspacing

\bibitem{msr09}
\BIBentryALTinterwordspacing
A.~Mockus, ``Amassing and indexing a large sample of version control systems:
  Towards the census of public source code history,'' in \emph{Proceedings of
  the 2009 6th IEEE International Working Conference on Mining Software
  Repositories}, ser. MSR '09.\hskip 1em plus 0.5em minus 0.4em\relax
  Washington, DC, USA: IEEE Computer Society, 2009, pp. 11--20. [Online].
  Available: \url{http://dx.doi.org/10.1109/MSR.2009.5069476}
\BIBentrySTDinterwordspacing

\bibitem{ma2016crowdsourcing}
Y.~Ma, T.~Dey, J.~M. Smith, N.~Wilder, and A.~Mockus, ``Crowdsourcing the
  discovery of software repositories in an educational environment,''
  \emph{PeerJ Preprints}, vol.~4, p. e2551v1, 2016.

\bibitem{greenfield2003software}
J.~Greenfield and K.~Short, ``Software factories: assembling applications with
  patterns, models, frameworks and tools,'' in \emph{Companion of the 18th
  annual ACM SIGPLAN conference on Object-oriented programming, systems,
  languages, and applications}.\hskip 1em plus 0.5em minus 0.4em\relax ACM,
  2003, pp. 16--27.

\bibitem{levy2003poisoning}
E.~Levy, ``Poisoning the software supply chain,'' \emph{Security \& Privacy,
  IEEE}, vol.~1, no.~3, pp. 70--73, 2003.

\bibitem{chacon2014pro}
S.~Chacon and B.~Straub, \emph{Pro Git}, 2nd~ed.\hskip 1em plus 0.5em minus
  0.4em\relax Berkely, CA, USA: Apress, 2014.

\bibitem{christopher1992}
M.~L. Christopher, \emph{Logistics and Supply Chain Management}.\hskip 1em plus
  0.5em minus 0.4em\relax London: Pitman Publishing, 1992.

\bibitem{chopra2007supply}
S.~Chopra and P.~Meindl, ``Supply chain management. strategy, planning \&
  operation,'' in \emph{Das Summa Summarum des Management}.\hskip 1em plus
  0.5em minus 0.4em\relax Springer, 2007, pp. 265--275.

\bibitem{John2011}
\BIBentryALTinterwordspacing
J.~T. Mentzer, W.~DeWitt, J.~S. Keebler, S.~Min, N.~W. Nix, C.~D. Smith, and
  Z.~G. Zacharia, ``Defining supply chain management,'' \emph{Journal of
  Business Logistics}, vol.~22, no.~2, pp. 1--25. [Online]. Available:
  \url{https://onlinelibrary.wiley.com/doi/abs/10.1002/j.2158-1592.2001.tb00001.x}
\BIBentrySTDinterwordspacing

\bibitem{Hau1997}
\BIBentryALTinterwordspacing
H.~L. Lee, V.~Padmanabhan, and S.~Whang, ``Information distortion in a supply
  chain: The bullwhip effect,'' \emph{Management Science}, vol.~43, no.~4, pp.
  546--558, 1997. [Online]. Available:
  \url{https://doi.org/10.1287/mnsc.43.4.546}
\BIBentrySTDinterwordspacing

\bibitem{buurman2002supply}
J.~Buurman, \emph{Supply chain logistics management}.\hskip 1em plus 0.5em
  minus 0.4em\relax McGraw-Hill, 2002.

\bibitem{maworld}
Y.~Ma, C.~Bogart, S.~Amreen, R.~Zaretzki, and A.~Mockus, ``World of code: An
  infrastructure for mining the universe of open source vcs data.''

\bibitem{survival}
\BIBentryALTinterwordspacing
T.~T, \emph{survival: A Package for Survival Analysis in S}, 2015, r package
  version 2.38. [Online]. Available:
  \url{https://CRAN.R-project.org/package=survival}
\BIBentrySTDinterwordspacing

\bibitem{mlogit}
\BIBentryALTinterwordspacing
Y.~Croissant, \emph{mlogit: multinomial logit model}, 2013, r package version
  0.2-4. [Online]. Available: \url{https://CRAN.R-project.org/package=mlogit}
\BIBentrySTDinterwordspacing

\bibitem{benjamin2018redefine}
D.~J. Benjamin, J.~O. Berger, M.~Johannesson, B.~A. Nosek, E.-J. Wagenmakers,
  R.~Berk, K.~A. Bollen, B.~Brembs, L.~Brown, C.~Camerer \emph{et~al.},
  ``Redefine statistical significance,'' \emph{Nature Human Behaviour}, vol.~2,
  no.~1, p.~6, 2018.

\bibitem{Bird:2006:MES:1137983.1138016}
\BIBentryALTinterwordspacing
C.~Bird, A.~Gourley, P.~Devanbu, M.~Gertz, and A.~Swaminathan, ``Mining email
  social networks,'' in \emph{Proceedings of the 2006 International Workshop on
  Mining Software Repositories}, ser. MSR '06.\hskip 1em plus 0.5em minus
  0.4em\relax New York, NY, USA: ACM, 2006, pp. 137--143. [Online]. Available:
  \url{http://doi.acm.org/10.1145/1137983.1138016}
\BIBentrySTDinterwordspacing

\bibitem{Mockus:2014:EBD:2593882.2593889}
\BIBentryALTinterwordspacing
A.~Mockus, ``Engineering big data solutions,'' in \emph{Proceedings of the on
  Future of Software Engineering}, ser. FOSE 2014.\hskip 1em plus 0.5em minus
  0.4em\relax New York, NY, USA: ACM, 2014, pp. 85--99. [Online]. Available:
  \url{http://doi.acm.org/10.1145/2593882.2593889}
\BIBentrySTDinterwordspacing

\bibitem{ms2007software}
------, ``Software support tools and experimental work,'' in \emph{Empirical
  Software Engineering Issues. Critical Assessment and Future
  Directions}.\hskip 1em plus 0.5em minus 0.4em\relax Springer, 2007, pp.
  91--99.

\bibitem{GBM16}
\BIBentryALTinterwordspacing
I.~Gorton, A.~B. Bener, and A.~Mockus, ``Software engineering for big data
  systems,'' \emph{IEEE Softw.}, vol.~33, no.~2, pp. 32--35, Mar. 2016.
  [Online]. Available: \url{http://dx.doi.org/10.1109/MS.2016.47}
\BIBentrySTDinterwordspacing

\bibitem{hora2015apiwave}
\BIBentryALTinterwordspacing
A.~Hora and M.~T. Valente, ``Apiwave: Keeping track of api popularity and
  migration,'' in \emph{Proceedings of the 2015 IEEE International Conference
  on Software Maintenance and Evolution (ICSME)}, ser. ICSME '15.\hskip 1em
  plus 0.5em minus 0.4em\relax Washington, DC, USA: IEEE Computer Society,
  2015, pp. 321--323. [Online]. Available:
  \url{http://dx.doi.org/10.1109/ICSM.2015.7332478}
\BIBentrySTDinterwordspacing

\bibitem{mileva2010mining}
\BIBentryALTinterwordspacing
Y.~M. Mileva, V.~Dallmeier, and A.~Zeller, ``Mining api popularity,'' in
  \emph{Proceedings of the 5th International Academic and Industrial Conference
  on Testing - Practice and Research Techniques}, ser. TAIC PART'10.\hskip 1em
  plus 0.5em minus 0.4em\relax Berlin, Heidelberg: Springer-Verlag, 2010, pp.
  173--180. [Online]. Available:
  \url{http://dl.acm.org/citation.cfm?id=1885930.1885952}
\BIBentrySTDinterwordspacing

\bibitem{Cossette:2012:SGT:2393596.2393661}
\BIBentryALTinterwordspacing
B.~E. Cossette and R.~J. Walker, ``Seeking the ground truth: A retroactive
  study on the evolution and migration of software libraries,'' in
  \emph{Proceedings of the ACM SIGSOFT 20th International Symposium on the
  Foundations of Software Engineering}, ser. FSE '12.\hskip 1em plus 0.5em
  minus 0.4em\relax New York, NY, USA: ACM, 2012, pp. 55:1--55:11. [Online].
  Available: \url{http://doi.acm.org/10.1145/2393596.2393661}
\BIBentrySTDinterwordspacing

\bibitem{Lammel:2011:LAA:1982185.1982471}
\BIBentryALTinterwordspacing
R.~L\"{a}mmel, E.~Pek, and J.~Starek, ``Large-scale, ast-based api-usage
  analysis of open-source java projects,'' in \emph{Proceedings of the 2011 ACM
  Symposium on Applied Computing}, ser. SAC '11.\hskip 1em plus 0.5em minus
  0.4em\relax New York, NY, USA: ACM, 2011, pp. 1317--1324. [Online].
  Available: \url{http://doi.acm.org/10.1145/1982185.1982471}
\BIBentrySTDinterwordspacing

\bibitem{Nguyen:2010:GAA:1932682.1869486}
\BIBentryALTinterwordspacing
H.~A. Nguyen, T.~T. Nguyen, G.~Wilson, Jr., A.~T. Nguyen, M.~Kim, and T.~N.
  Nguyen, ``A graph-based approach to api usage adaptation,'' \emph{SIGPLAN
  Not.}, vol.~45, no.~10, pp. 302--321, Oct. 2010. [Online]. Available:
  \url{http://doi.acm.org/10.1145/1932682.1869486}
\BIBentrySTDinterwordspacing

\bibitem{kabinna2016logging}
\BIBentryALTinterwordspacing
S.~Kabinna, C.-P. Bezemer, W.~Shang, and A.~E. Hassan, ``Logging library
  migrations: A case study for the apache software foundation projects,'' in
  \emph{Proceedings of the 13th International Conference on Mining Software
  Repositories}, ser. MSR '16.\hskip 1em plus 0.5em minus 0.4em\relax New York,
  NY, USA: ACM, 2016, pp. 154--164. [Online]. Available:
  \url{http://doi.acm.org/10.1145/2901739.2901769}
\BIBentrySTDinterwordspacing

\bibitem{teyton2014study}
\BIBentryALTinterwordspacing
C.~Teyton, J.-R. Falleri, M.~Palyart, and X.~Blanc, ``A study of library
  migrations in java,'' \emph{Journal of Software: Evolution and Process},
  vol.~26, no.~11, pp. 1030--1052. [Online]. Available:
  \url{https://onlinelibrary.wiley.com/doi/abs/10.1002/smr.1660}
\BIBentrySTDinterwordspacing

\bibitem{de2018library}
\BIBentryALTinterwordspacing
F.~L. de~la Mora and S.~Nadi, ``Which library should i use?: A metric-based
  comparison of software libraries,'' in \emph{Proceedings of the 40th
  International Conference on Software Engineering: New Ideas and Emerging
  Results}, ser. ICSE-NIER '18.\hskip 1em plus 0.5em minus 0.4em\relax New
  York, NY, USA: ACM, 2018, pp. 37--40. [Online]. Available:
  \url{http://doi.acm.org/10.1145/3183399.3183418}
\BIBentrySTDinterwordspacing

\bibitem{teyton2012mining}
C.~Teyton, J.-R. Falleri, and X.~Blanc, ``Mining library migration graphs,'' in
  \emph{19th Working Conference on Reverse Engineering, WCRE 2012, Kingston,
  ON, Canada, October 15-18, 2012}, 2012, pp. 289--298.

\bibitem{Tonelli2010}
T.~T. Bartolomei, K.~Czarnecki, R.~L{\"a}mmel, and T.~van~der Storm, ``Study of
  an api migration for two xml apis,'' in \emph{Software Language Engineering},
  M.~van~den Brand, D.~Ga{\v{s}}evi{\'{c}}, and J.~Gray, Eds.\hskip 1em plus
  0.5em minus 0.4em\relax Berlin, Heidelberg: Springer Berlin Heidelberg, 2010,
  pp. 42--61.

\bibitem{Tonelli2011}
T.~Tonelli, Krzysztof, and Ralf, ``Swing to swt and back: Patterns for api
  migration by wrapping,'' in \emph{2010 IEEE International Conference on
  Software Maintenance}, Sept 2010, pp. 1--10.

\bibitem{Dagenais2009}
B.~Dagenais and M.~P. Robillard, ``Semdiff: Analysis and recommendation support
  for api evolution,'' in \emph{2009 IEEE 31st International Conference on
  Software Engineering}, May 2009, pp. 599--602.

\bibitem{Mileva:2009:MTL:1595808.1595821}
\BIBentryALTinterwordspacing
Y.~M. Mileva, V.~Dallmeier, M.~Burger, and A.~Zeller, ``Mining trends of
  library usage,'' in \emph{Proceedings of the Joint International and Annual
  ERCIM Workshops on Principles of Software Evolution (IWPSE) and Software
  Evolution (Evol) Workshops}, ser. IWPSE-Evol '09.\hskip 1em plus 0.5em minus
  0.4em\relax New York, NY, USA: ACM, 2009, pp. 57--62. [Online]. Available:
  \url{http://doi.acm.org/10.1145/1595808.1595821}
\BIBentrySTDinterwordspacing

\bibitem{anderson1990choice}
E.~E. Anderson, ``Choice models for the evaluation and selection of software
  packages,'' \emph{Journal of Management Information Systems}, vol.~6, no.~4,
  pp. 123--138, 1990.

\bibitem{xiao2014social}
S.~Xiao, J.~Witschey, and E.~Murphy-Hill, ``Social influences on secure
  development tool adoption: why security tools spread,'' in \emph{Proceedings
  of the 17th ACM conference on Computer supported cooperative work \& social
  computing}.\hskip 1em plus 0.5em minus 0.4em\relax ACM, 2014, pp. 1095--1106.

\end{thebibliography}
%



\end{document}